\documentclass[showpacs,showkeys,longbibliography]{revtex4-1}

\usepackage{amssymb}
\usepackage{amsmath}
\usepackage{graphicx}
\usepackage{amsbsy}
\usepackage{float}
\usepackage{color}

\newcommand{\bq}{\begin{eqnarray}}
\newcommand{\eq}{\end{eqnarray}}
\newcommand{\bqn}{\begin{eqnarray*}}
\newcommand{\eqn}{\end{eqnarray*}}
\newcommand{\nn}{{\bf n}}
\newcommand{\rr}{{\bf r}}

\newcommand{\an}[1]{#1}

\begin{document}
\title{A cluster theory for a Janus fluid} 

\author{Riccardo Fantoni}
\email{rfantoni27@sun.ac.za}
\affiliation{National Institute for Theoretical Physics (NITheP) and
Institute of Theoretical Physics, Stellenbosch University,
Stellenbosch 7600, South Africa} 

\date{\today}

\begin{abstract}
Recent Monte Carlo simulations on the Kern and Frenkel model of a Janus
fluid have revealed that in the vapour phase there is the formation of
preferred clusters made up of a well-defined number of particles:
the micelles and the vesicles. A cluster theory is developed to
approximate the exact clustering properties stemming from the
simulations. It is shown that the theory is able to reproduce the
micellisation phenomenon.  
\end{abstract}


\maketitle
\section{Introduction}
\label{sec:introduction}

\an{In the statistical mechanics of fluids \cite{Hill} the liquid
state \cite{Hansen} is a particularly fascinating one. A liquid is neither
a gas nor a solid, but the state where correlations really play an
important role. The pioneering work of Berni J. Alder \cite{Alder57}
showed that, because of the absence of attractive forces, the
hard-sphere fluid  admits only a single fluid phase. In order to find the
liquid phase it is sufficient to add an attractive square-well to the
pair-potential of the hard-spheres. The resulting hard-sphere
square-well fluid admits a bell-shaped gas-liquid coexistence curve 
\cite{Vega1992,*Liu2005} with the critical point moving at low
temperatures and high densities as the attractive well width
diminishes. Recently N. Kern and D. Frenkel \cite{Kern03} studied,
through computer experiments, a new fluid model made of hard-spheres
with patchy square-well attractions. In its simplest version, the {\sl
single patch} case, the model only depends on the {\sl surface
coverage} $\chi$ of the patch and the attraction range. Between the
two extreme cases $\chi=0$, the hard-sphere model, and $\chi=1$, the
hard-sphere square-well model, where the particles pair-potential is
isotropic, the particles interaction is directional. The $\chi=1/2$
model is known as the Janus case, as the particle, like the roman God,
has two faces of different functionalities. 

Another important process, which may lead to the manifestation of
macroscopic phenomena, in certain fluids, is the clustering
or association. In 1956, for example, Leon N. Cooper \cite{Cooper1956}
found that the stable state of the degenerate electron fluid in
a metal is one in which particles of opposite spin and opposite momentum
form pairs. It was then understood that whereas the electrons in a metal
form pairs with relative angular momentum zero, in $\mbox{}^3$He this
would be prevented by the hard core repulsion, and that therefore
Cooper pairing had to occur in a state of finite angular momentum. In
1961 A. Lenard \cite{Lenard1961} proved analytically that a 
two-component plasma living in one dimension undergoes a transition
from the conducting to the insulating state by the formation of
neutral dimers made of a positive and a negative charge. A
two-component plasma living in two dimensions is only stable at 
sufficiently high temperatures \cite{Hauge71}. But if one adds a hard
core to the charges it remains stable even at low temperatures where
it undergoes the same transition \cite{Kosterlitz1973}. The hard core
gives rise to anyonic statistics for the quantum fluid living in two
dimensions \cite{Lerda}. In three dimensions the two-component plasma
with a hard core, the so called restricted-primitive model, also
undergoes the clustering transition at low temperature and low
densities \cite{Valeriani2010}. An example of a one-component Janus
fluid undergoing association is the dipolar hard-sphere fluid. Here a
particle can be viewed as the superposition of two uniformly charged
spheres: a positive one and a negative one \cite{Rovigatti2011}.}

In their study of the Kern and Frenkel single patch $\chi=1/2$
Janus case, F. Sciortino {\sl et al.} \cite{Sciortino2009} found that
the gas branch of the coexistence curve bends at high densities at low
temperatures. Below the critical point, the fluid 
tends to remain in the gas phase for a larger interval of
densities respect to the $\chi=1$ case. This behaviour is due to the
tendency of particles to associate due to the directional attractive
component in the pair-potential and 
form clusters. At low temperatures, these clusters interact weakly
amongst themselves because the particles of which they are composed
tend to expose the hard-sphere hemisphere on the outside of the
collapsed cluster.   

By studying the clustering properties of the gas phase of the Janus
fluid, F. Sciortino {\sl et al.} discovered that below the critical
temperature there is a range of temperatures where there is 
formation of two kinds of preferred clusters: the {\sl micelles} and
the {\sl vesicles}. In the former the particles tend to arrange
themselves into a spherical shell and in the latter they tend to
arrange themselves as two concentric spherical shells.
      
\an{It is important to confront existing cluster theories with these new
findings based on computer experiments. In this work the Bjerrum
cluster theory for electrolytes, later extended by A. Tani
\cite{Tani83} to include trimers, has been employed (preliminary
results appeared in Ref. \onlinecite{Fantoni2011}) for the description
of the exact equilibrium cluster concentrations found in the computer
experiment of F. Sciortino {\sl et al.}. The theory is extended to   
clusters of up to $12$ particles in an attempt to reproduce the
micellisation phenomenon observed in the simulations around a reduced
temperature of $0.27$. A different determination of the intra-cluster
configurational partition function has been devised in place of the
one used by J. K. Lee \cite{Lee1973}.}   

\an{The Kern and Frenkel fluid has been used to describe} soft matter
\cite{deGennes1992} \an{biological and non-biological materials like
globular proteins in solution \cite{Kern03,Giacometti2010,Romano2010}
and colloidal suspensions 
\cite{Kern03,Romano2011}, or molecular liquids
\cite{Romano2011b}}. \an{Recently} there has been a  
tremendous development in the techniques for the synthesis of patchy 
colloidal particles \cite{Glotzer2007,*Pawar2010} in the
laboratory. These are particles with dimensions of $10-
10^4$\AA$\;$ in diameter, which obey to Boltzmann statistics
\an{\cite{stat}}. From the realm of patchy colloidal particles stems
the family of Janus 
particles for their simplicity \cite{Casagrande1989,*Walther2008}. It
is possible to create Janus particles in the laboratory in large
quantities \cite{Hong2006} and to study their clustering properties
\cite{Hong2006b,*Hong2008}.  

The micelles and the vesicles are complex structures observed in the
chemistry of surfactant molecules \an{analogous to those which may be
found in the physical biology of the cell \cite{Phyllips}}.

The paper is organized as follows: in Section \ref{sec:model} we describe
the fluid model, in Section \ref{sec:clustering-properties} we
present the clustering properties of the fluid found in the Monte
Carlo simulations of F. Sciortino {\sl et al.}, the cluster theory is
presented and 
developed in Sections \ref{sec:cluster-theory} and \ref{sec:cpf}, in
Section \ref{sec:results} we compare the numerical results from our
approximation to the exact results of F. Sciortino {\sl et al.}, and Section
\ref{sec:conclusions} is for final remarks.   

\section{The Kern and Frenkel model}
\label{sec:model}

As in the work of F. Sciortino {\sl et al.} \cite{Sciortino2009} we use
the Kern and Frenkel \cite{Kern03} single patch hard-sphere model of the
Janus fluid. Two spherical particles attract via a square-well 
potential only if the line joining the centers of the two spheres
intercepts the patch on the surfaces of both particles. The
pair-potential is separated as follows  
\bq
\Phi(1,2)=\phi(r_{12})\Psi(\hat{\nn}_1,\hat{\nn}_2,\hat{\rr}_{12})~,
\eq
where
\bq
\phi(r)=\left\{
\begin{array}{ll}
+\infty   & r<\sigma\\
-\epsilon & \sigma<r<\lambda\sigma\\
0         & \lambda\sigma<r
\end{array}\right.
\eq
and
\bq
\Psi(\hat{\nn}_1,\hat{\nn}_2,\hat{\rr}_{12})=\left\{
\begin{array}{ll}
1 & \mbox{if $\hat{\nn}_1\cdot\hat{\rr}_{12}\ge\cos\theta_0$ and 
$-\hat{\nn}_2\cdot\hat{\rr}_{12}\ge\cos\theta_0$}\\
0 & \mbox{otherwise}
\end{array}\right.
\eq
where $\theta_0$ is the angular semi-amplitude of the patch. Here
$\hat{\nn}_i(\omega_i)$ are versors pointing from the center of sphere
$i$ to the center of the attractive patch, with
$\omega_i$ their solid angles and 
$\hat{\rr}_{12}(\Omega)$ is the versor pointing from the center of
sphere 1 to the center of sphere 2, with $\Omega$ its solid angle. 
We denote with $\sigma$ the hard core diameter and
$\lambda=1+\Delta/\sigma$ with $\Delta$ the width of the attractive
well.

A particle configuration is determined by its position and its
orientation. 

We will use $\sigma$ as the unit of length and $\epsilon$ as the unit
of energy.

One can determine the fraction of the particle surface covered by the
attractive patch as follows  
\bq
\chi=\langle\Psi(\hat{\nn}_1,\hat{\nn}_2,\hat{\rr}_{12})
\rangle_{\omega_1,\omega_2}^{1/2}
=\sin^2\left(\frac{\theta_0}{2}\right)~,
\eq  
where $\langle \ldots \rangle_{\omega}=\int \ldots d\omega/(4\pi)$. 

As in the work of F. Sciortino {\sl et al.} \cite{Sciortino2009} we limit
ourselves to the {\sl Janus case} $\chi=1/2$.

\section{Clustering properties}
\label{sec:clustering-properties}

The Janus fluid just described will undergo clustering as there is
a directional attractive component in the interaction between its
particles. Moreover at low temperatures the collapsed clusters are
expected to interact weakly with each other. This is responsible
for the bending at high density of the low temperature gas branch
of the gas-liquid binodal curve recently determined in
Ref. \onlinecite{Sciortino2009}. \an{Below the critical temperature,
in the vapour phase, the appearance of weakly interacting clusters
destabilizes the liquid phase in favour of the gas phase}. F. Sciortino
{\sl et al.} during their canonical ensemble (at fixed number of
particles $N$, volume $V$, and  temperature $T$, with $\rho=N/V$ the
density) Monte Carlo simulations 
of the fluid also studied its clustering properties. In particular
they used the following topological definition of a cluster: an
ensemble of $n$ particles form a {\sl cluster} when, 
starting from one particle, is possible to reach all other particles
through a path. The path being allowed to move from one particle to
another if there is attraction between the two particles. During the
simulation of the fluid they counted the number $N_n$ of clusters of $n$
particles, which depends on the particles configurations, and took a
statistical average of this number. 

\begin{figure}[H]
\begin{center}
\includegraphics[width=12cm]{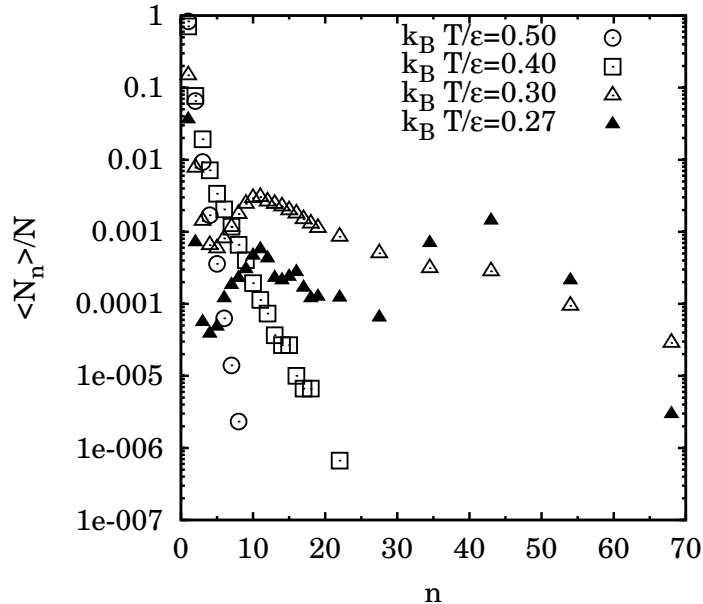}
\end{center}
\caption{Exact cluster concentrations of the Janus fluid with
  $\Delta=\sigma/2$ at a reduced density $\rho\sigma^3=0.01$ and various
  reduced temperatures $k_BT/\epsilon$, from the Monte Carlo
  simulation of F. Sciortino {\sl et al.} \cite{Sciortino2009}.}
\label{fig:cp}
\end{figure}

We show in Fig. \ref{fig:cp} the results they obtained for
$\Delta=\sigma/2$ at a reduced density $\rho\sigma^3=0.01$ and various
reduced temperatures $k_BT/\epsilon$. From the figure we can see how
at a reduced temperature of $0.27$, in the vapour phase, there is the
formation of two kinds of preferred clusters: one made up of around $10$
particles and one made up of around $40$ particles.  

In their collapsed shape, expected at low temperatures, the particles
in the clusters tend to expose their inactive hemisphere 
on the outside of the cluster, resulting in a weak interaction between
pairs of clusters.

In the clusters of around $10$ particles the particles tend to arrange
themselves into a spherical shell, forming a micellar structure. In
the clusters of around $40$ particles the particles are arranged into
two concentric spherical shells, forming a vesicular structure.  

The aim of the present work is to see if we can approximate the exact
equilibrium cluster concentrations found in the simulation using a
cluster theory. We will restrict ourselves to clusters made of up to
$12$ particles to see if the theory is able to reproduce the
micellisation phenomenon. The theory is described next.

\section{A cluster Theory for Janus particles}
\label{sec:cluster-theory}
Following Ref. \onlinecite{Tani83}, we describe the fluid of $N$
particles undergoing clustering as a mixture of $N$ species of
clusters. Clusters of species $n=1,\ldots,N$, which we call
$n$-clusters, are made up of $n$ 
particles. We denote with $N_n$ the number of clusters of species $n$
and with $\rho_n=N_n/V$ their density. We assume that the chemical
potentials of all the cluster species are zero (there is no cost in
energy in the formation or destruction of a cluster). Then the
grand-canonical partition function of the fluid can be written as   
\bq
\label{cluster:eq1}
Q_{\text{tot}}&=&\sum_{\{N_n\}}^{\prime} \prod_{n=1}^{N}
\frac{1}{N_n!} \left(q_n^{\text{intra}}\right)^{N_n} Q_{\text{inter}}
\left(\{N_n\},V,T\right)~, 
\eq
where one separates the coordinates and momenta relative to the center
of mass of a cluster from the ones of the center of mass so that
$q_n^{\text{intra}}$ will be the intra-cluster partition function of
the cluster of species $n$ and $Q_{\text{inter}}$ the inter-cluster
partition function where we consider the clusters as non identical. 
The prime indicates that the sum is restricted by the condition that
the number of particles of the fluid is $N$,
\bq
\label{cluster:eq2}
\sum_{n=1}^N n N_n &=& N~.
\eq

We approximate $Q_\text{tot}$ assuming that
the sum can be replaced by its largest dominant contribution. Using
the Stirling approximation $N!\approx (N/e)^N$ one then obtains 
\bq
\label{cluster:eq4}
\ln Q_{\text{tot}}&\approx& \sum_{n=1}^N \left[N_n \ln
q_n^{\text{intra}} - \left(N_n \ln N_n - N_n\right)\right] + \ln
Q_{\text{inter}}~.
\eq
The maximum of $\ln Q_\text{tot}$ as a function of $\{N_n\}$ on the
constraint of Eq. (\ref{cluster:eq2}) is given by the point
$\{\overline{N}_n\}$ where the gradients of $\ln Q_\text{tot}$ and 
of the constraint have the same direction. Introducing a Lagrange
multiplier $\lambda$ the equilibrium cluster distribution
$\{\overline{N}_n\}$ is then found from the conditions
\bq
\label{cluster:eq5}
\left.\frac{\partial}{\partial N_n} \ln Q_{\text{tot}}
\right|_{\{N_n=\overline{N}_n\}} +\ln\lambda^n &=& 0~,~~~n=1,2,3,\ldots 
\eq
The resulting Helmholtz free energy, $\beta F_{\text{tot}}=-\ln
Q_{\text{tot}}$, can then be written in terms of the intra-cluster 
free energy, $\beta f_{n}^{\text{intra}}=-\ln q_n^{\text{intra}}$,
and the inter-cluster partition function as follows
\bq
\label{cluster:eq6}
\frac{\beta F_{\text{tot}}}{V} &=& \sum_{n=1}^{N}
\left[\overline{\rho}_n \ln \overline{\rho}_n - \overline{\rho}_n
\right] + 
\sum_{n=1}^N \overline{\rho}_n \beta f_{n}^{\text{intra}} +
\sum_{n=1}^N \overline{\rho}_n \ln V - \frac{1}{V} \ln
Q_{\text{inter}}~.
\eq
where $\beta=1/k_BT$ with $k_B$ Boltzmann constant and
$\overline{\rho}_n=\overline{N}_n/V$. 

We expect the equilibrium cluster concentrations, $\overline{N}_n/N$, to
approximate the ones measured in the simulation, $\langle N_n\rangle/N$.
   
\section{Relationship between the configurational partition functions}
\label{sec:cpf}

We will assume that Eq. (\ref{cluster:eq1}) also holds at the level
of the configurational partition functions $Z$, as follows 
\bq
\label{cluster:eq7}
Z_{\text{tot}}&=&\sum_{\{N_n\}}^{\prime} \prod_{n=1}^{N}
\frac{1}{N_n!} \left(z_n^{\text{intra}}\right)^{N_n} Z_{\text{inter}}
\left(\{N_n\},V,T\right)~. 
\eq
In the calculation we only work at the level of the configurational
partition functions. 

Since we expect the clusters to be weakly interacting amongst themselves
we will approximate the inter-clusters configurational partition
function with: {\sl i.} the ideal gas approximation for pointwise
clusters and {\sl ii.} the Carnahan-Starling approximation
\cite{Carnahan69} for clusters of diameter $\sigma_0$. A third
possibility, that we have not investigated, would be to use the
Boubl\'ik, Mansoori, Carnahan, and Starling approximation
\cite{Boublik70,Mansoori71} for clusters of different diameters
$\sigma_n$.  

We will only work with a limited number $n_c$ of different cluster
species. Since we are investigating whether the cluster
theory is able to reproduce the micellisation phenomenon we will only
consider the first $n_c$ clusters: $n=1,2,3,\ldots,n_c$. And choosing
$n_c=12$.  

We will describe next the two approximations used for the inter-cluster
configurational partition function.

\subsection{Ideal gas approximation}
\label{subsec:ig}

The simplest possibility is to approximate the mixture of clusters
as an ideal one so that  
\bq
Z_{inter}=V^{N_t}~,
\eq 
where $N_t=\sum_n N_n$ is the total number of clusters.

The equations for the equilibrium numbers of clusters are 
\bq \label{ni}
\overline{N}_n&=&\lambda^n V z_n^{\text{intra}}~,~~~n=1,2,3,\ldots,n_c\\ 
N&=&\sum_n n\overline{N}_n~,
\eq
from which we can determine all the concentrations $\overline{N}_n/N$
and the Lagrange multiplier by solving the resulting algebraic
equation of order $n_c$. The case $n_c=2$ is described in Appendix
\ref{app:wertheim}.
 
\subsection{Carnahan-Starling approximation}
\label{subsec:cs}

A better approximation is found if we use as the inter-cluster
configurational partition function the Carnahan-Starling expression
\cite{Carnahan69} for hard-spheres of diameter $\sigma_0$,
\bq
\ln Z_{inter}=N_t\ln V-N_t\frac{\eta_t(4-3\eta_t)}{(1-\eta_t)^2}~,
\eq
where $\eta_t=(\pi/6)\rho_t\sigma_0^3$ is the clusters packing
fraction and $\rho_t=N_t/V$ their density. 

In this case one needs to solve a system of $n_c+1$ coupled transcendental
equations,  
\bq
\overline{N}_n&=&\lambda^n V
z_n^{\text{intra}}G(\overline{\eta}_t)~,~~~i=1,2,3,\ldots,n_c\\  
N&=&\sum_n n\overline{N}_n~,
\eq
with $\overline{\eta}_t=(\pi/6)\overline{\rho}_t\sigma_0^3$,
$\overline{\rho}_t=\overline{N}_t/V$, $\overline{N}_t=\sum_n
\overline{N}_n$, and   
\bq
G(x)=\exp\left[-\frac{x(8-9x+3x^2)}{(1-x)^3}\right]~.
\eq

In order to search for the correct root of this system of equations it
is important to choose the one that is continuously obtained from the
physical solution of the ideal gas approximation as $\sigma_0\to 0$.
\an{Giving a volume to the clusters we introduce correlations between
them which will prove to be essential for a qualitative reproduction of
the micellisation phenomenon though the cluster theory. The
Carnahan-Starling approximation amounts to choosing for the sequence of
virial coefficients of the hard-spheres, a general term which is a
particular second order polynomial and to determine the polynomial
coefficients that approximate the third virial coefficient by its
closest integer \cite{Carnahan69}. It could be interesting to repeat
the calculation using for the inter-cluster partition function the
hard-spheres one choosing all but the first virial coefficient equal
to zero, to see if that is sufficient to reproduce the micellisation
phenomena.} 

Note that in order to study the vesicles we would have to solve a
system of around 40 coupled equations.

We will describe next how do we determine the intra-cluster
configurational partition function $z_n^{\text{intra}}$.

\subsection{The intra-cluster configurational partition function}
\label{sec:ic}

To estimate the intra-cluster configurational partition function we
performed Monte Carlo simulations of an {\sl isolated} topological
cluster. 

We determined the reduced excess internal energy per particle of the
$n$-cluster $u_n^{ex}=\langle \sum_{i<j}^n\Phi(i,j)\rangle/(n\epsilon)$
($u_1^{ex}=0$ by definition) as a function of the temperature, and
then used thermodynamic integration to determine the intra-cluster
configurational partition function.  
 
We found that the results for $u_n^{ex}(T^\star)$ can be fitted by a
Gaussian as follows 
\bq \label{ugauss}
u_n^{ex}(T^\star)=a_ne^{-b_n{T^\star}^2}+c_n~,
\eq
with $T^\star=k_BT/\epsilon$ the reduced temperature.

Given the excess free energy of the $n$-cluster
$F_n^{ex,\text{intra}}$, we can then determine 
$f^{ex,\text{intra}}_{n}=\beta F^{ex,\text{intra}}_{n}/n$ as follows 
\bq \nonumber
f^{ex,\text{intra}}_{n}(\beta^\star)&=&\int_0^{\beta^\star} u_n^{ex}(1/x)\,dx\\
&=& c_n\beta^\star+a_n\sqrt{b_n}\left\{
\frac{e^{-b_n/{\beta^\star}^2}}{\sqrt{b_n/{\beta^\star}^2}}+
\sqrt{\pi}\left[\mbox{erf}\left(\sqrt{b_n/{\beta^\star}^2}\right)
-1\right]\right\}~,
\eq 
with $\beta^\star=1/T^\star$ and $v_0=\pi\sigma_0^3/6$ the volume of
the $n$-cluster. Then the
intra-cluster configurational partition function is given by
$z_n^{\text{intra}}=v_0^n\exp(-nf^{ex,\text{intra}}_{n})$ with
$z_1^{\text{intra}}=v_0$. 

We studied only the first $10$ clusters with $n=3,\ldots,12$. The
dimer being trivial.
To this end we started with an initial configuration of two
pentagons with particles at their vertexes juxtaposed one above the
other. The two pentagons are parallel to the $(x,y)$ plane, have the
$z$ axis passing through their centers, and are placed one at
$z=+\sigma/2$ and 
the other at $z=-\sigma/2$. The particles patches all point towards the
origin. We formed the clusters with a lower number of particles by
simply deleting particles and the clusters with 11 and 12 particles by
adding a particle on the $z$ axis just above the upper pentagon and
just below the lower one. 

\begin{figure}[H]
\begin{center}
\includegraphics[width=12cm]{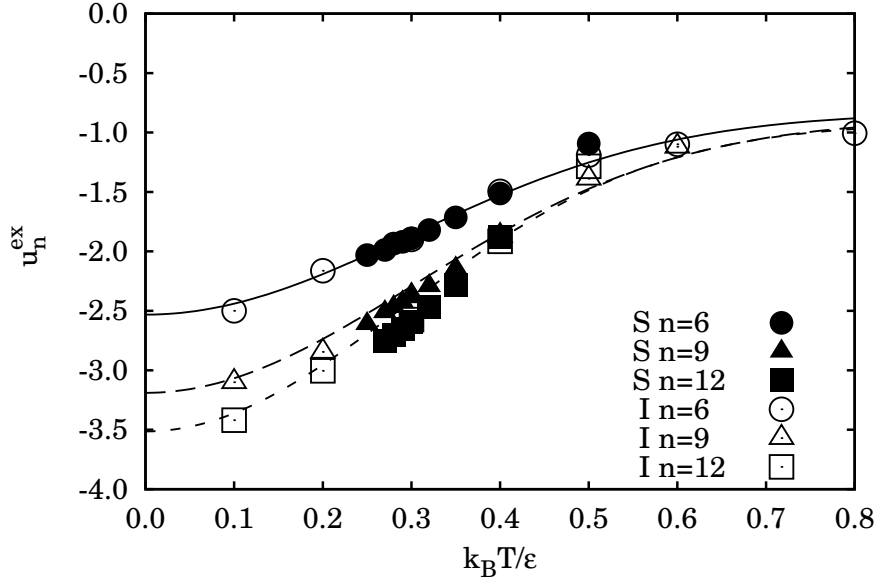}
\end{center}
\caption{Reduced excess internal energy per particle as a function of
  temperature for the 6-, 9-, and 12-cluster. The results from the
  isolated (I) cluster calculation are compared with the results of
  Sciortino (S) for the Janus fluid with $\Delta=\sigma/2$ at a
  reduced density $\rho\sigma^3=0.01$. Also shown is the Gaussian fit
  of Eq. (\ref{ugauss}).}  
\label{fig:ic-fc}
\end{figure}

We performed the simulations of the isolated cluster at a fictitious
reduced density of $\rho\sigma^3=0.05$ which ensured a simulation box big
enough that the cluster did not percolate through the periodic
boundary conditions. We also compared our results for the excess
internal energy calculation for the isolated cluster with the results of
F. Sciortino {\sl et al.} for the low density Janus fluid, from which
one extracts 
cluster information by taking all the clusters found with 
the same number of particles and averaging their properties, as shown
in Fig. \ref{fig:ic-fc}. 

At high temperatures the limiting value for the excess internal energy
per particle of the isolated $n$-cluster is $-\epsilon (n-1)/n$ corresponding
to the stretched cluster. At low temperature ($T^\star<0.15$) the cluster
tends to freeze into certain energy minima. So in order to reach the
absolute minimum we used the following smoothing procedure. We smoothed
the Kern and Frenkel potential by choosing  
\bq
\Psi(\hat{\nn}_1,\hat{\nn}_2,\hat{\rr}_{12})=
\{\tanh[l(\hat{\nn}_1\cdot\hat{\rr}_{12}-\cos\theta_0)]+1\}
\{\tanh[l(-\hat{\nn}_2\cdot\hat{\rr}_{12}-\cos\theta_0)]+1\}/4~.
\eq  
We then gradually changed the parameter $l$, during the simulation,
starting from $1/2$ and increasing up to values where there is no
actual difference between the smoothed potential and the original
stepwise one.  
The reduced excess internal energy per particle and gyration radii for
such minimum energy configurations are shown in Table 
\ref{tab:2}. 

\begin{table}[H]\scriptsize
\begin{center}
\begin{tabular}{|cccc|}
\hline
$n$ & $\langle U\rangle/(\epsilon n)$ & $\langle U\rangle/\epsilon$ & $R_g$\\
\hline
1       &0      &0      & 0\\
2       &-0.5   &-1     & $\sim 1/2$\\
3       &-1     &-3     & $\sim 1/\sqrt{3}$\\
4       &-1.5   &-6     & 0.83\\
5       &-2.0   &-10    & 0.76\\
6       &-2.50  &-15    & 0.75\\
7       &-2.71  &-19    & 0.91\\
8      	&-2.88	&-23    & 0.93\\
9	&-3.10	&-28    & 0.96\\
10	&-3.20	&-32    & 1.00\\
11	&-3.36	&-37    & 1.04\\
12	&-3.42	&-41    & 1.08\\
\hline
\end{tabular}
\end{center}
\caption{The low temperature reduced excess internal energy per particle
$\langle U\rangle/(\epsilon n)$ ($U$ is the potential energy of the
cluster) of 
the clusters with up to 12 particles when $\Delta=\sigma/2$. Also shown
is the gyration radius $R_g^2=\sum_{i=1}^n|\rr_i-\rr_{cm}|^2/n$ with
$\rr_{cm}=\sum_{i=1}^n\rr_i/n$, $\rr_i$ being the position of the
$i$-th particle in the cluster.} 
\label{tab:2}
\end{table}

In the Metropolis algorithm \cite{Kalos-Whitlock} used to sample the
probability distribution function proportional to $e^{-\beta U}$,
where $U$ is the potential energy of the cluster, the random walk
moves through the configuration space of the particles forming the
cluster through two kinds of moves: a displacement of the  
particle position and a rotation of the particle (through the
Marsaglia algorithm \cite{Allen}). We followed two different
strategies in the simulations: {\sl i.} we averaged only over the
particles configurations that form a cluster and {\sl ii.} we explicitly
modified the acceptance probability by rejecting moves that break the
cluster. So in the second strategy all the moves are counted in
the averages. The two strategies turned out to give the same results,
as they should. The second strategy is
preferable to simulate the bigger clusters at high temperature and
for small well widths because there is no loss of statistics.

In Appendix \ref{app:tab} we present the results for the reduced
excess internal energy of the isolated clusters as a function of
temperature and their fit of Eq. (\ref{ugauss}).

\subsection{Thermodynamic quantities}
\label{sec:thermodynamics}
Once the equilibrium cluster distribution $\{\overline{N}_n\}$ has been
determined (within the ideal gas or the Carnahan-Starling
approximation for the inter-cluster partition function) the
configurational partition function $Z_\text{tot}$ is 
known. Then the excess free energy is  
\bq
\beta F^{ex}=-\ln\left(\frac{Z_\text{tot}}{V^N}\right)~,
\eq
the reduced internal energy per particle of the fluid is
\bq
u=\frac{3}{2\beta^\star}+
\frac{1}{N}\frac{\partial(\beta F^{ex})}{\partial\beta^\star}
=\frac{3}{2\beta^\star}-\sum_n\frac{\overline{N}_n}{N}
\frac{\partial\ln z_n^{\text{intra}}}{\partial\beta^\star}
=\frac{3}{2\beta^\star}+\sum_n n\frac{\overline{N}_n}{N}u_n^{ex}~,
\eq
and its compressibility factor, in the Carnahan-Starling approximation
for the inter-cluster configurational partition function, is
\bq
\frac{\beta P}{\rho}=
\frac{1}{\rho}\frac{\partial\ln Z_\text{tot}}{\partial V}\approx
\frac{1}{\overline{\rho}_t}\frac{\partial\ln Z_\text{inter}}{\partial V}=
\frac{1+\overline{\eta}_t+\overline{\eta}_t^2-\overline{\eta}_t^3}
{(1-\overline{\eta}_t)^3}~.
\eq
\an{Here we have used the approximation $N\approx\overline{N}_t$
which turns out to be reasonable at the chosen value of the cluster
diameter, as shown in Fig. \ref{fig:ts0}.} 

In Fig. \ref{fig:thermo} we show the results for the compressibility
factor and the reduced excess internal energy per particle. The
reduced excess internal energy is compared with the Monte Carlo data
of F. Sciortino {\sl et al.} (Fig. 1 in Ref. \onlinecite{Sciortino2009}). 

\section{Results}
\label{sec:results}

We present here the numerical results from the cluster theory and
compare them with the results of F. Sciortino {\sl et al.} from the
simulation of the Janus ($\chi=1/2$) fluid with $\Delta=\sigma/2$. 

We studied three different attraction ranges: $\Delta=\sigma/2$,
$\Delta=\sigma/4$, and $\Delta=0.15\sigma$. To the best of our
knowledge there are no Monte Carlo results available for the two
smaller ranges. 

We only present the results obtained from the Carnahan-Starling
approximation for the inter-cluster partition function as the ideal
gas approximation turned out to be too crude an approximation even for
a qualitative description of the exact clustering properties.

\subsection{$\Delta=\sigma/2$}

For $\Delta=\sigma/2$ we found the following results.

\subsubsection{Equilibrium cluster concentrations}

\begin{figure}[H]
\begin{center}
\includegraphics[width=12cm]{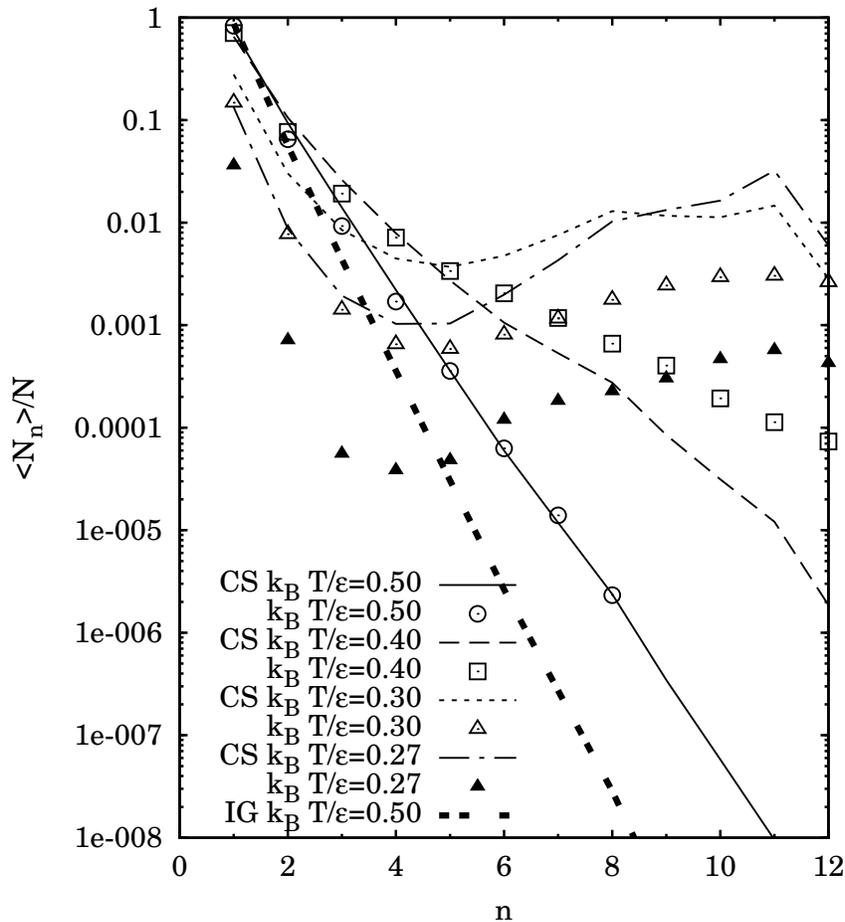}
\end{center}
\caption{Comparison between the Monte Carlo (MC) data (points) and the
Carnahan-Starling 
(CS) approximation with $\sigma_0=2.64\sigma$ (lines) for the cluster
concentrations $\langle N_n\rangle/N$, $n=1,2,3,\ldots,12$,
as a function of the cluster size $n$ at $\rho\sigma^3=0.01$ and various
temperatures. Also shown is the ideal gas (IG) approximation at the
same density and the highest temperature $k_BT/\epsilon=0.5$.}    
\label{fig:NNs}
\end{figure}

\begin{figure}[H]
\begin{center}
\includegraphics[width=12cm]{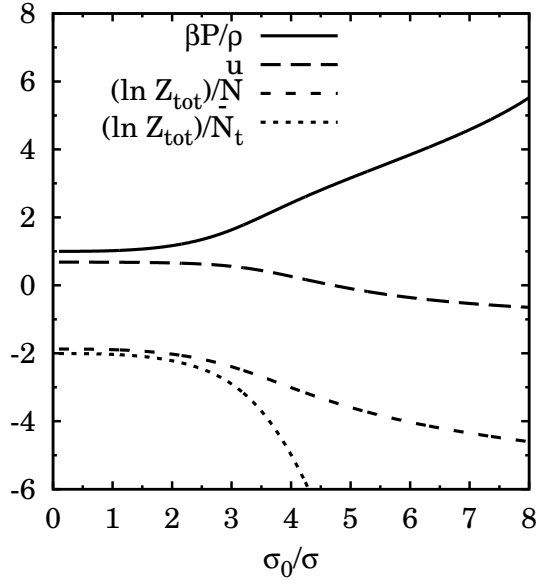}
\end{center}
\caption{The compressibility factor, the internal energy per particle,
  and the logarithm of the total partition function per total number
  of particles and per total number of clusters as a function of the
  clusters diameter $\sigma_0$ at the thermodynamic state
  $\rho\sigma^3= 0.01$ and $k_BT/\epsilon= 0.5$ for $\Delta=0.5\sigma$.}    
\label{fig:ts0}
\end{figure}

In Fig. \ref{fig:NNs} we compare the Monte Carlo data of F. Sciortino
{\sl et al.} (the results reported in Fig. \ref{fig:cp}) and our results 
from the cluster theory. From the figure one can see that the ideal gas
approximation for the inter-cluster partition function is not
appropriate even at high temperatures in the single fluid phase above
the critical point. In order to find agreement with the Monte Carlo
data at high 
temperatures it is sufficient to give a volume to the clusters,
treating them as hard-spheres of a diameter $\sigma_0$. In the
Carnahan-Starling approximation we gradually increased $\sigma_0$ from
zero and found that for $\sigma_0=2.64\sigma$ the results of the
cluster theory were in good agreement with the Monte Carlo data at
$k_BT/\epsilon=0.5$. Using the same 
cluster diameter at all other temperatures, we saw that the theory
is able to qualitatively reproduce the micellisation phenomenon
observed in the simulation of F. Sciortino {\sl et al.}.  
The results also suggest that with a temperature-dependent cluster
diameter, or more generally with a cluster diameter dependent on the
thermodynamic state of the fluid, we could achieve better agreement
between our approximation and the exact results. \an{Our topological
definition of a cluster has no direct geometrical
interpretation. Other definitions with a geometrical nature are
possible. For example Lee {\sl et al.} in their studies of nucleation
define an assembly of particles to be a cluster if they all lie within
a sphere of radius $\sigma_0$ centered on one of the particles. In
our simulations of the isolated clusters these have a globular shape
at low temperature and a necklace shape at high temperature. The
optimal cluster diameter $\sigma_0=2.64\sigma$ (found to give good
agreement between the exact and approximate clusters concentrations at
high temperature) suggests necklace clusters made up of around 3 particles
or globular clusters made up of around $2\pi(\sigma_0/\sigma)^2/\sqrt{3}
\approx 25$ particles placed on a spherical shell. Since $\sigma_0$
is the only free parameter of the theory, it is important to estimate how
thermodynamic quantities like the compressibility factor $\beta
P/\rho$, the reduced internal energy per particle $u$, and the
logarithm of the total configurational partition function per number
of particles, $\ln Z_{tot}/N$, or per number of clusters, $\ln
Z_{tot}/\overline{N}_t$, are sensible to variations in $\sigma_0$. From
Fig. \ref{fig:ts0} we can see that for the thermodynamic state
$\rho\sigma^3=0.01$ and $k_BT/\epsilon=0.5$, the thermodynamic
quantities are roughly independent of $\sigma_0$ for
$\sigma_0\lesssim 3\sigma$.}

\begin{figure}[H]
\begin{center}
\includegraphics[width=12cm]{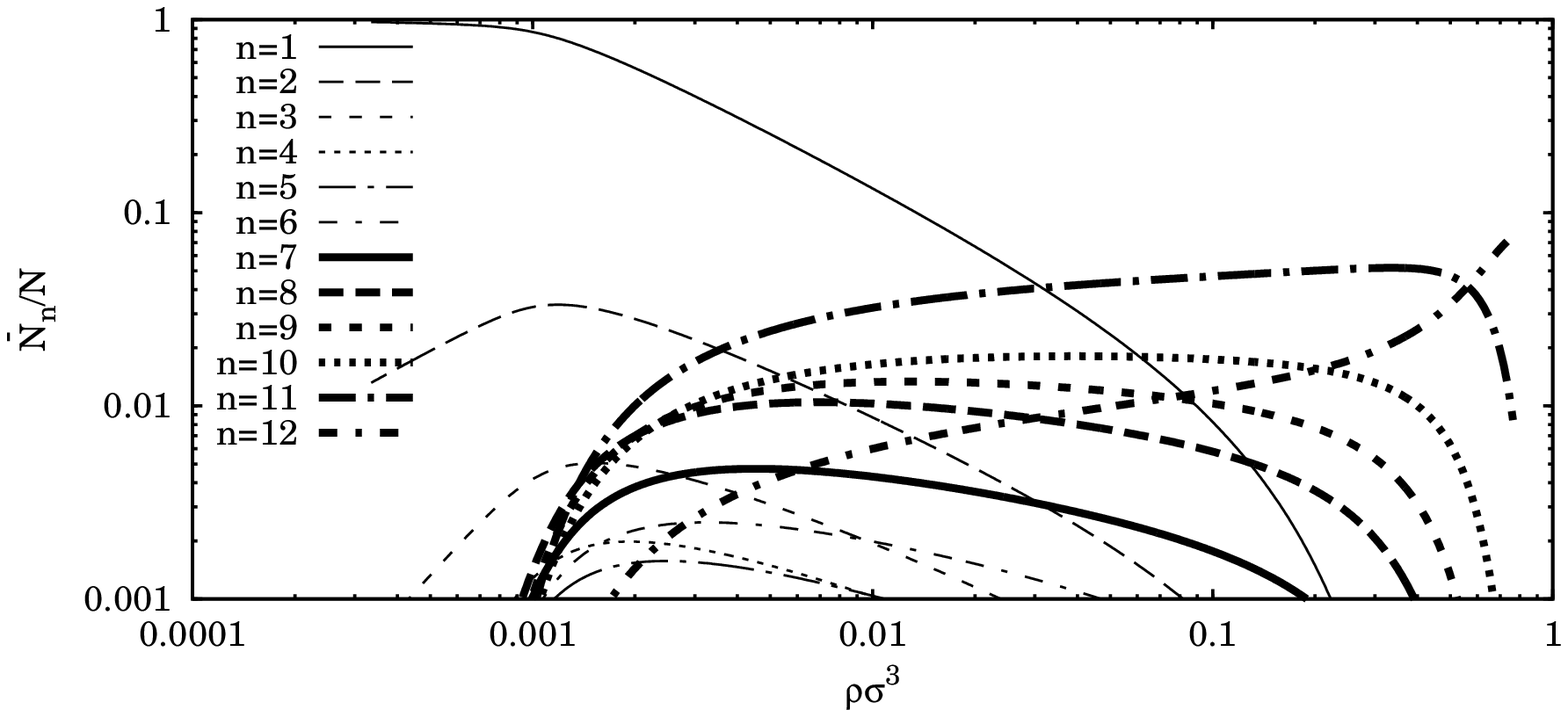}\\
\includegraphics[width=12cm]{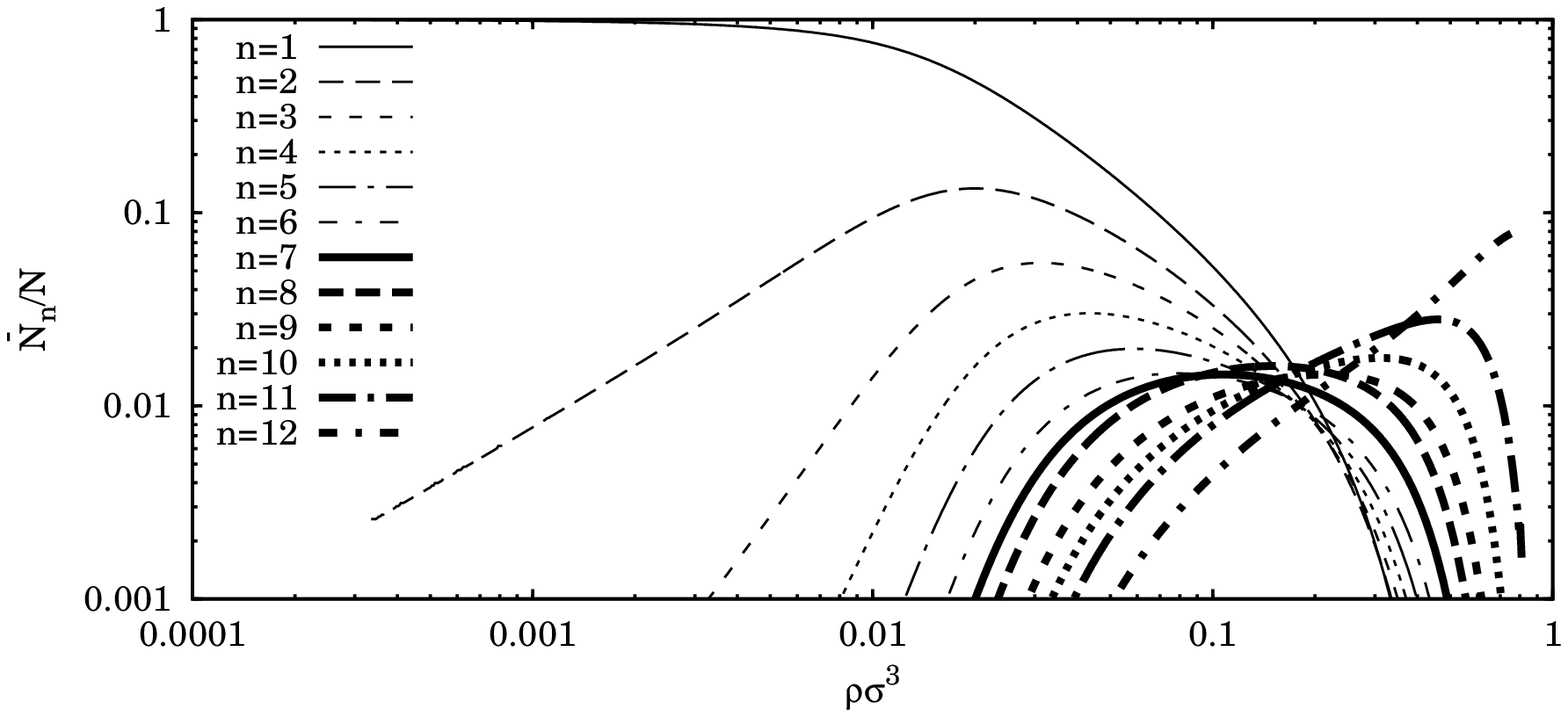}
\end{center}
\caption{The equilibrium cluster concentrations $\overline{N}_n/N$,
  $n=1,2,3,\ldots,12$, as a function of density for
  $k_BT/\epsilon=0.27$ (top panel) and $k_BT/\epsilon=0.5$ (bottom
  panel) as obtained from the CS approximation with
  $\sigma_0=2.64\sigma$. Here $\Delta=\sigma/2$.}    
\label{fig:nval-cs-0.5}
\end{figure}

In Fig. \ref{fig:nval-cs-0.5} we show the behaviour of the equilibrium
cluster concentrations, from the Carnahan-Starling approximation with
$\sigma_0=2.64\sigma$, as a function of density at $k_BT/\epsilon=0.27$.
From the figure we can see that at very low densities there are
essentially no clusters. But as the density increases, clusters of an
increasing number of particles appear in the fluid. In particular, at
$k_BT/\epsilon=0.27$ there is an interval of densities where 
clusters of $11$ particles are preferred.

\subsubsection{Thermodynamic quantities}

\begin{figure}[H]
\begin{center}
\includegraphics[width=12cm]{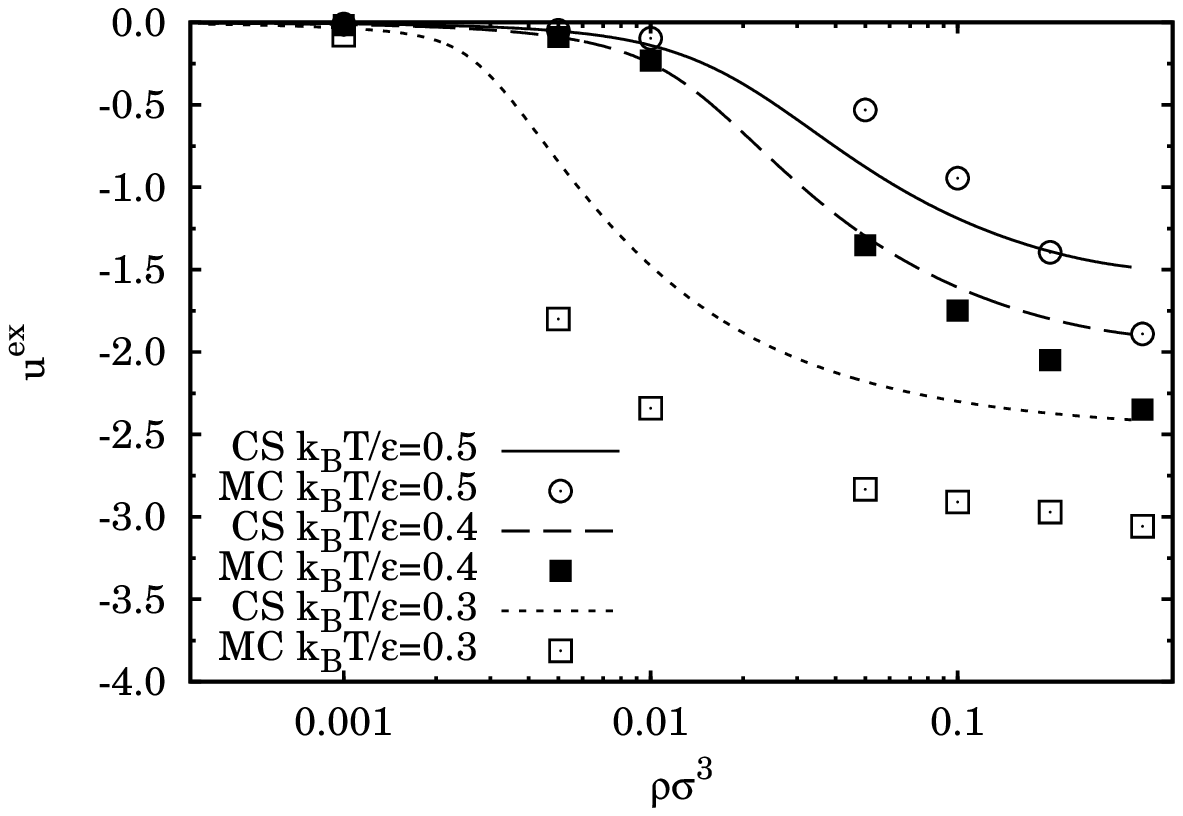}\\
\includegraphics[width=12cm]{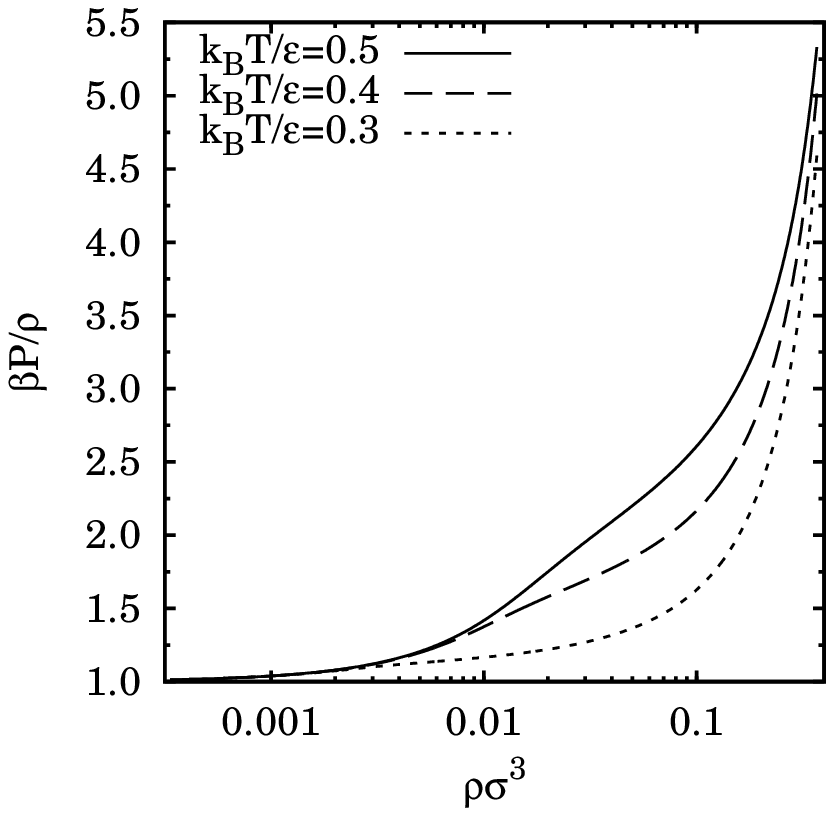}
\end{center}
\caption{The top panel shows the reduced excess internal energy per particle
  for three different values of temperature as a function of
  density. The results from the Carnahan-Starling (CS) approximation
  are compared with the Monte Carlo (MC) results of F. Sciortino {\sl
  et al.} \cite{Sciortino2009}. The bottom panel shows the 
  compressibility factor for the same values of temperature as a
  function of density from the CS approximation (no MC data is
  available).}    
\label{fig:thermo}
\end{figure}

Following Section \ref{sec:thermodynamics} we now use the cluster
theory within the Carnahan-Starling approximation with
$\sigma_0=2.64\sigma$ to extract thermodynamic information for the
Janus fluid. In Fig. \ref{fig:thermo} we show the results 
obtained for the excess reduced internal energy per particle and the
compressibility factor. 

From the figure we see that there is a qualitative agreement between
the results of the cluster theory and the Monte Carlo results. No
Monte Carlo results are available for the compressibility factor.

\subsection{$\Delta=\sigma/4$}

\begin{figure}[H]
\begin{center}
\includegraphics[width=12cm]{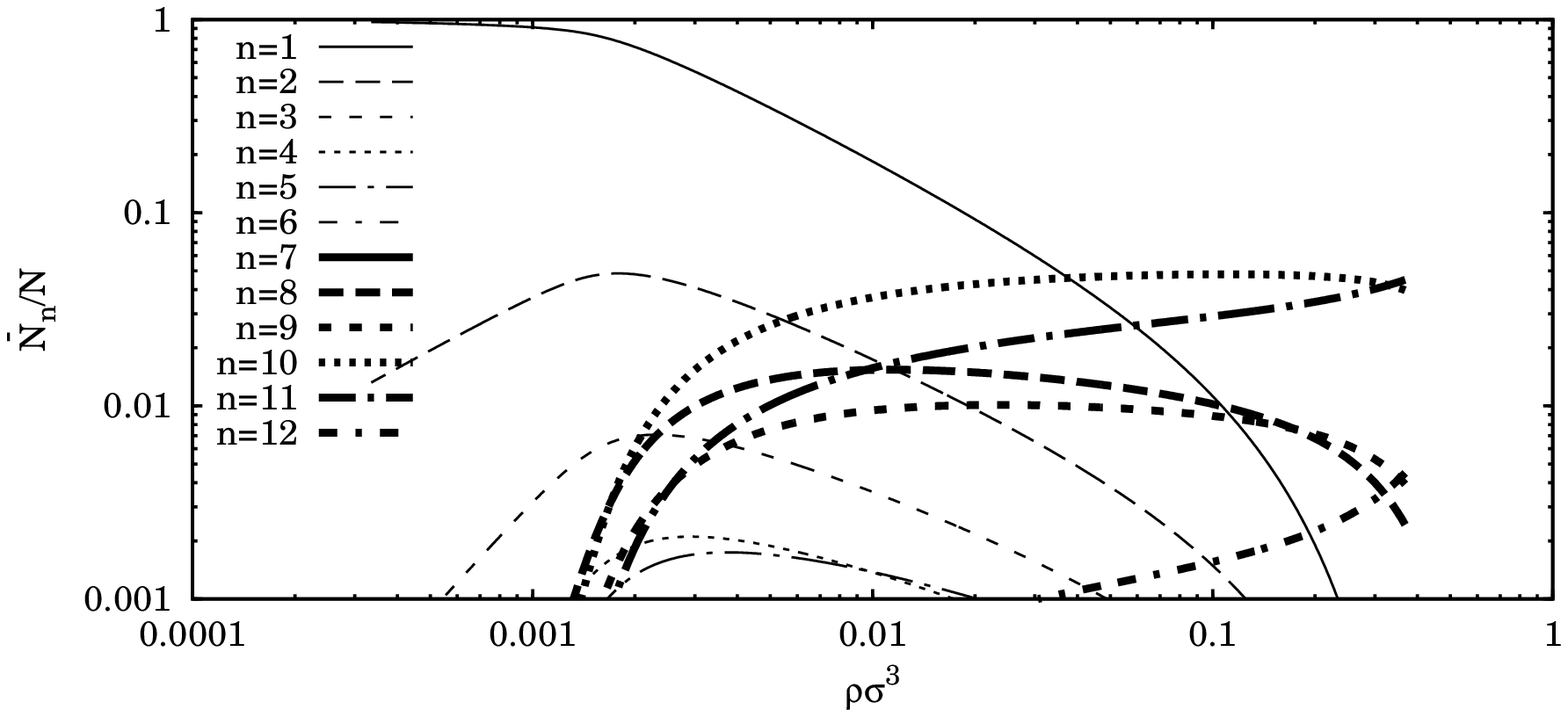}\\
\includegraphics[width=12cm]{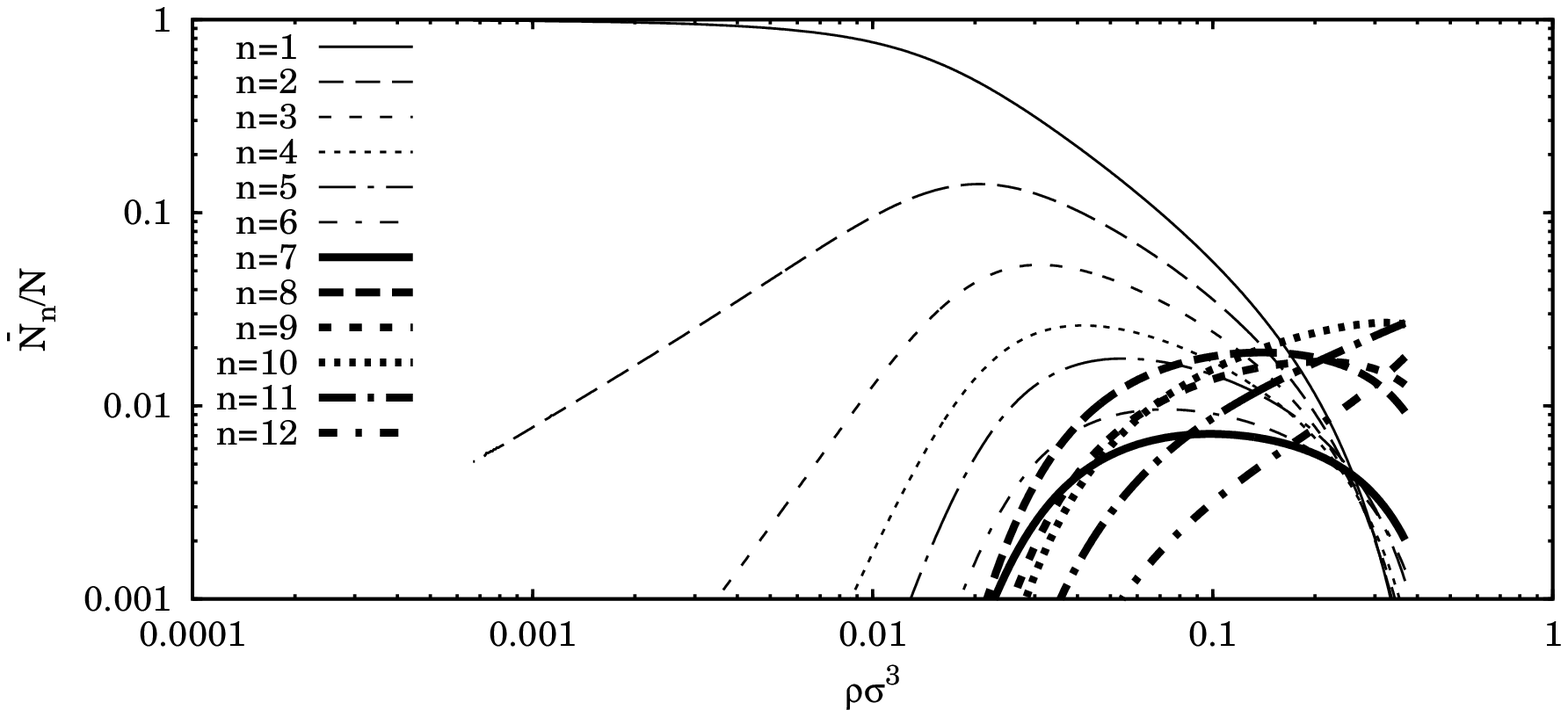}
\end{center}
\caption{Same as Fig. \ref{fig:nval-cs-0.5} for $\Delta=\sigma/4$.}    
\label{fig:nval-cs-0.25}
\end{figure}

Decreasing the width of the attractive well to $\Delta=\sigma/4$ yielded
the results shown in Fig. \ref{fig:nval-cs-0.25}. We see that
now, at the reduced temperature $0.27$, the preferred clusters are the
ones made up of $10$ particles. 
%

\subsection{$\Delta=0.15\sigma$}

\begin{figure}[H]
\begin{center}
\includegraphics[width=12cm]{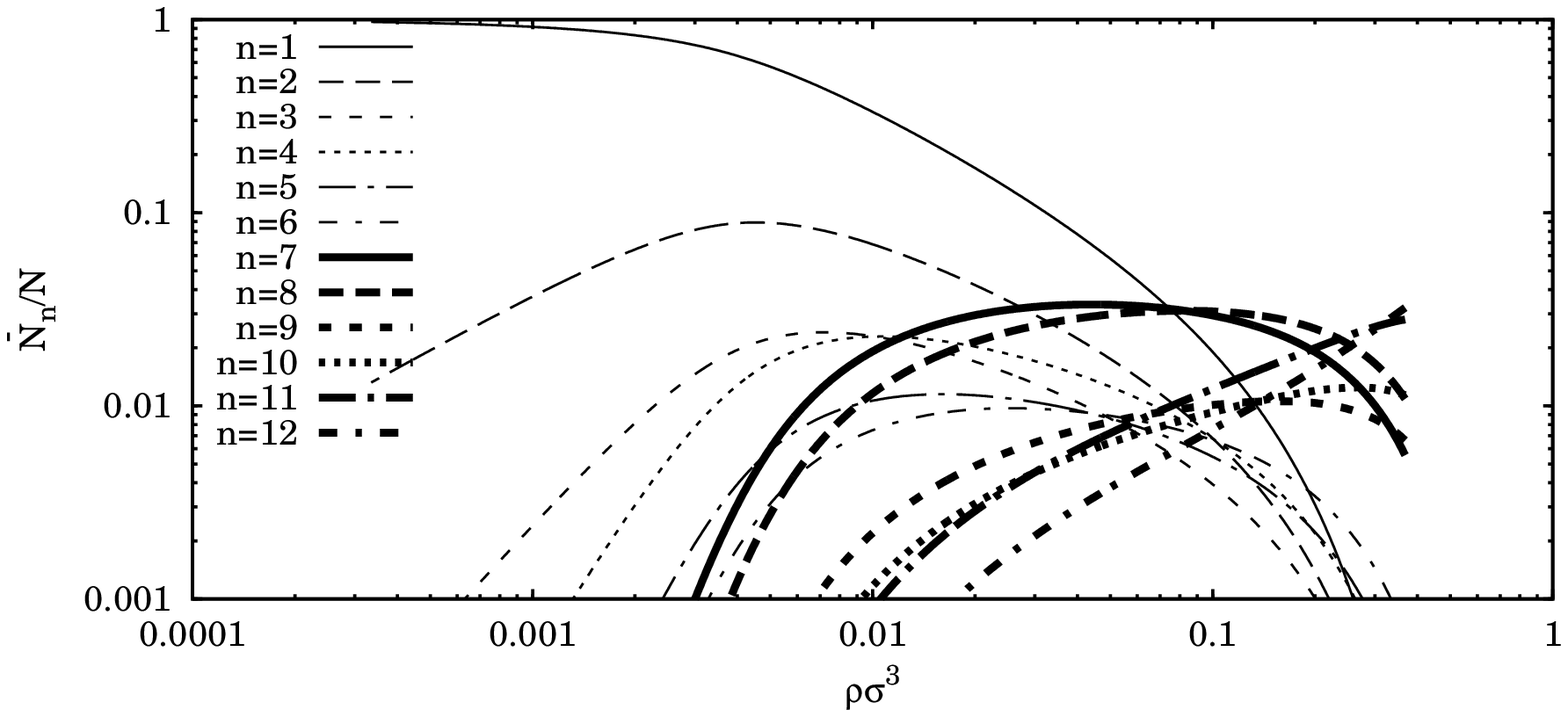}\\
\includegraphics[width=12cm]{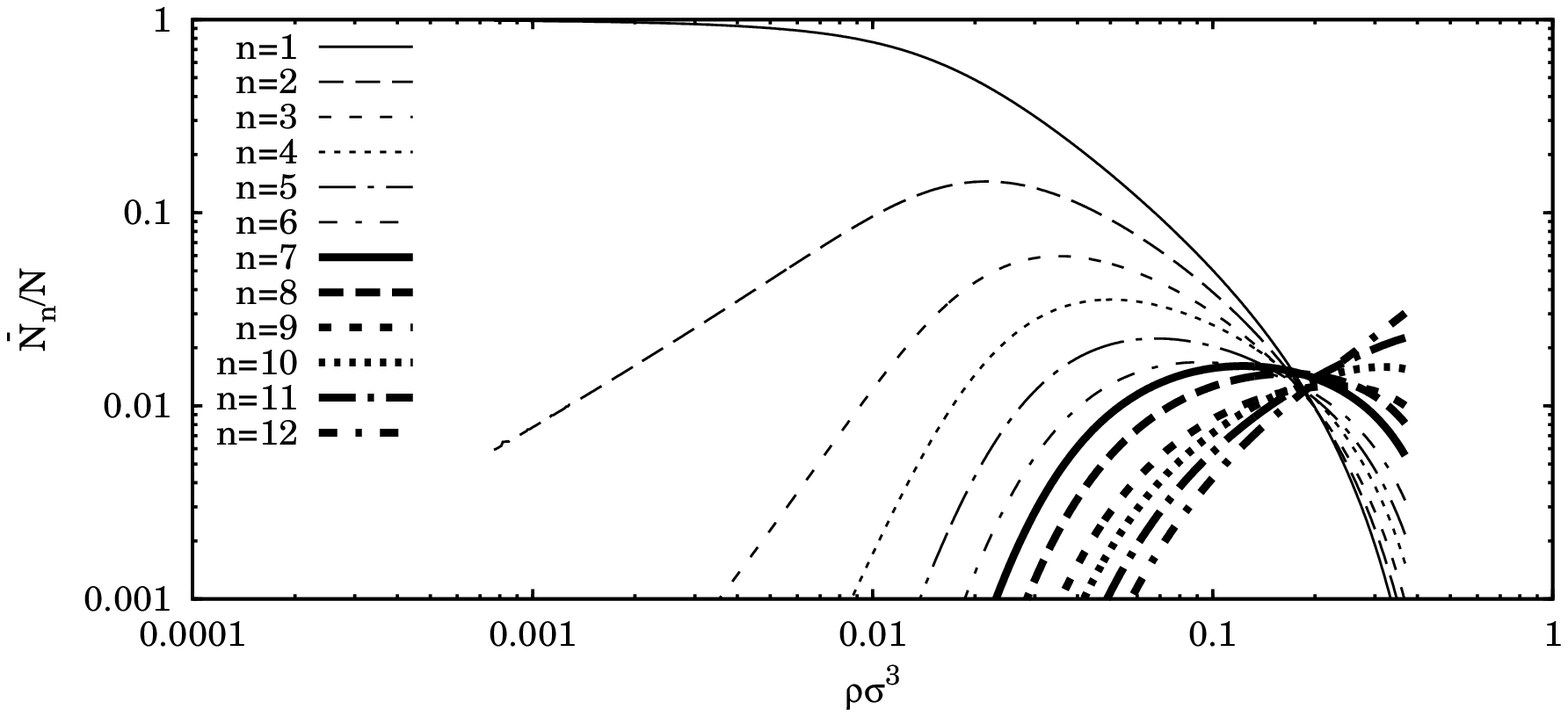}
\end{center}
\caption{Same as Fig. \ref{fig:nval-cs-0.5} for $\Delta=0.15\sigma$.}
\label{fig:nval-cs-0.15}
\end{figure}

Decreasing the width of the attractive well even further to
$\Delta=0.15\sigma$, we obtained the results of
Fig. \ref{fig:nval-cs-0.15}. Now, at the reduced temperature $0.27$,
\an{there is a range of densities around $\rho\sigma^3=0.1$ where the
preferred clusters are made up of 7 or 8 particles.}
%

\section{Conclusions}
\label{sec:conclusions}

We constructed a cluster theory for a fluid undergoing clustering and
showed that it is able to reproduce the micellisation phenomena
recently observed in the simulation of the vapour phase of Kern
and Frenkel Janus particles \cite{Sciortino2009}. A topological
definition of the cluster is used. We determined the intra-cluster
configurational partition function through thermodynamic integration 
of the excess internal energy of the cluster, estimated
through Monte Carlo simulations of an isolated cluster. In the simulation
we restricted the random walk through the configurations of the particles
that compose the cluster by rejecting the moves that break the cluster. Due
to the geometrical characteristics of the pair-potential it is
expected that the clusters, when in their collapsed shape, will be very weakly 
interacting amongst themselves as the Janus particles will expose the
hard-sphere hemisphere on the outside of the cluster. We thus used
for the estimation of the inter-cluster configurational partition
function first the simple ideal gas approximation for pointwise clusters
and then the Carnahan-Starling approximation for clusters seen as
hard-spheres of diameter $\sigma_0$. The equilibrium cluster
concentrations obtained with the ideal gas approximation
turned out to disagree, even at high temperatures, with the ones 
obtained from the simulation of the fluid \cite{Sciortino2009} and
were not able to reproduce the micellisation phenomenon in the vapour
phase. We then 
gradually increased $\sigma_0$ from zero until we found good agreement
between the equilibrium cluster concentrations obtained with the
Carnahan-Starling approximation and the concentrations from the simulation of
the fluid \cite{Sciortino2009} at high temperature (above the critical
point). Using the same value of $\sigma_0$ for lower temperatures
(below the critical point) we were able to qualitatively reproduce
the micellisation phenomenon observed in the simulation of the fluid
\cite{Sciortino2009} around a reduced temperature of $0.27$ and 
a reduced density of $0.01$. \an{This result is important for two
reasons. Firstly it shows that the clustering fundamentally arising
from the canonical ensemble description of the fluid of particles
can be approximated by a grand canonical ensemble description
of a particular clustered fluid. Secondly the second description,
which assumes from the start a clustered structure of the fluid, is much
less computationally costly than the first. Unlike most previous
works on cluster theories where the aim is usually to avoid the
Monte Carlo simulation \cite{Tani83,Caillol95}, our approach is a
hybrid one where we still 
use the Monte Carlo experiment to determine the intra-cluster
properties. Of course our goal can only be a qualitative description of
the fluid as we specifically prescribe a particular description of the
clusters and this is the source of our approximation.}  

Studying the behaviour of the equilibrium cluster concentrations as a
function of density and temperature, we saw that the micellisation
phenomenon only takes place within a particular range of temperatures
(below the critical point) and densities (in the vapour phase).

Once the equilibrium concentrations have been found it is possible
to determine how the cluster theory approximates the thermodynamic
quantities of the fluid. 
We find qualitative agreement between the Monte Carlo data of
F. Sciortino {\sl et al.} \cite{Sciortino2009} and our approximation
for the excess 
internal energy of the vapour phase. For the compressibility factor
no Monte Carlo data is available \an{so our results remain a theoretical
prediction.} 

We studied three different values of the attractive square-well width:
$\Delta=\sigma/2$, $\Delta=\sigma/4$, and $\Delta=0.15\sigma$. Monte
Carlo results \cite{Sciortino2009} are available only for the largest
width. Our study shows that as the range of the attraction diminishes
the micelles tend to be made up of a smaller number of particles. 

A related interesting problem to that just discussed is the one of
trying to give a definition of a liquid drop expected to form in the
coexistence region as a result of the condensation instability.  

\appendix
\section{Connection with Wertheim association theory}
\label{app:wertheim}

At small $\chi$, allowing only clusters of one (monomers) and two
(dimers) particles, we get 
\bq
\overline{N}_1&=&\lambda V z_1^\text{intra}~,\\
\overline{N}_2&=&\lambda^2 V z_2^\text{intra}~,\\
N&=&\overline{N}_1+2\overline{N}_2~,
\eq
which is a quadratic equation in $\lambda$. The solution for the 
fraction of patches that are not bonded (fraction of monomers) is 
\bq
\frac{\overline{\rho}_1}{\rho}=\frac{2}{1+\sqrt{1+8\rho\overline{\Delta}}}~,
\eq 
with
$\overline{\Delta}=z_2^{\text{intra}}/[z_1^{\text{intra}}]^2$ and
$\rho=N/V$ the density of the fluid, 
in accord, at low $T$, with the recent analysis of F. Sciortino {\sl
et al.} \cite{Sciortino2007} (compare their $X$ of Eq. (10) with our
$\overline{\rho}_1/\rho$ and their $\Delta$ with our
$\overline{\Delta}$), based on Wertheim association theory   
\cite{Wertheim-ass}. Our theory, contrary to the one of Wertheim,
allows to consider the case of multiple bonding of the patch. 

At high temperature our $\overline{\Delta}$ differs from the $\Delta$
of Ref. \onlinecite{Sciortino2007} but in this limit the clusters begin
to dissociate.

\section{Tables for the excess internal energy per particle of the clusters}
\label{app:tab}
We present here the results for the reduced excess internal energy per
particle as a function of temperature of the isolated $n$-cluster with
$n=2,3,\ldots,12$ as obtained from our Monte Carlo simulations.  

In Tab. \ref{tab:1} we show the results at $\Delta=0.5\sigma$ 
obtained with the strategy {\sl i.} described in Sec. \ref{sec:ic}. 
The smoothing procedure described in Sec. \ref{sec:ic} was used at the
lowest temperature. The excess internal energy per particle of the
$n=2$ cluster is always $-\epsilon/2$ given our topological definition
of a cluster. 

In Tab. \ref{tab:1.5} we show the results at $\Delta=0.5\sigma$
obtained with strategy {\sl ii.} described in Sec. \ref{sec:ic}. The
smoothing procedure described in Sec. \ref{sec:ic} was not used at the 
lowest temperature. Comparing Tabs. \ref{tab:1} and \ref{tab:1.5} we
can see that the two strategies lead to the same results.

In Tab. \ref{tab:4} we show the results at $\Delta=0.25\sigma$
obtained with strategy {\sl i.} described in Sec. \ref{sec:ic}. The
smoothing procedure described in Sec. \ref{sec:ic} was not used at the 
lowest temperature.

In Tab. \ref{tab:6} we show the results at $\Delta=0.15\sigma$
obtained with strategy {\sl ii.} described in Sec. \ref{sec:ic}. The
smoothing procedure described in Sec. \ref{sec:ic} was not used at the 
lowest temperature.

In Tab. \ref{tab:fit} we give the fit to the Gaussian of
Eq. (\ref{ugauss}) of the reduced excess internal energy per particle
as a function of the temperature. 

\begin{table}[H]\tiny
\begin{tabular}{|ccc|}
\hline
$k_BT/\epsilon$ & $\langle U\rangle/(\epsilon n)$ & error \\
\hline
$\infty$&-0.666   	&0\\
0.8&	-0.724&	        0.001\\
0.6&	-0.747& 	0.001\\
0.5&	-0.769&  	0.002\\
0.4&	-0.807& 	0.001\\
0.3&	-0.877& 	0.001\\
0.2&	-0.9663&	0.0008\\
0.1&	-1	&	$\approx$0\\
\hline 
\end{tabular}
\begin{tabular}{|ccc|}
\hline
$k_BT/\epsilon$ & $\langle U\rangle/(\epsilon n)$ & error \\
\hline
$\infty$&-0.75   	&0\\
0.8&	-0.849&	        0.004\\
0.6&	-0.898& 	0.004\\
0.5&	-0.961&  	0.005\\
0.4&	-1.081&	        0.004\\
0.3&	-1.278& 	0.003\\
0.2&	-1.460& 	0.002\\
0.1&	-1.5	&	$\approx$0\\
\hline 
\end{tabular}
\begin{tabular}{|ccc|}
\hline
$k_BT/\epsilon$ & $\langle U\rangle/(\epsilon n)$ & error \\
\hline
$\infty$&-0.8   	&0\\
0.8&	-0.942&	        0.009\\
0.6&	-0.995& 	0.008\\
0.5&	-1.085&	        0.008\\
0.4&	-1.322&	        0.007\\
0.3&	-1.606& 	0.004\\
0.2&	-1.792&	        0.003\\
0.1&	-2.0	&	$\approx$0\\
\hline 
\end{tabular}
\begin{tabular}{|ccc|}
\hline
$k_BT/\epsilon$ & $\langle U\rangle/(\epsilon n)$ & error \\
\hline
$\infty$&-0.833	&       0\\
0.8&	-1.01  &	0.03\\
0.6&	-1.10 & 	0.02\\
0.5&	-1.19 &	        0.01\\
0.4&	-1.49 &	        0.01\\
0.3&	-1.899 & 	0.009\\
0.2&	-2.16 & 	0.01\\
0.1&	-2.5	&	0\\
\hline 
\end{tabular}
\begin{tabular}{|ccc|}
\hline
$k_BT/\epsilon$ & $\langle U\rangle/(\epsilon n)$ & error \\
\hline
$\infty$&-0.857	&0\\
0.8&	-1.04& 	0.04\\
0.6&	-1.12& 	0.02\\
0.5&	-1.28&	0.02\\
0.4&	-1.68&	0.02\\
0.3&	-2.11& 	0.04\\
0.2&	-2.39& 	0.03\\
0.1&	-2.7	&	$\approx$0\\
\hline 
\end{tabular}
\begin{tabular}{|ccc|}
\hline
$k_BT/\epsilon$ & $\langle U\rangle/(\epsilon n)$ & error \\
\hline
$\infty$&-0.875		&0\\
0.8&	-1.06&	0.05\\
0.6&	-1.25&	0.05\\
0.5&	-1.27&	0.02\\
0.4&	-1.82&	0.02\\
0.3&	-2.26&	0.01\\
0.2&	-2.60&	0.02\\
0.1&	-2.9	&	$\approx$0\\
\hline 
\end{tabular}
\begin{tabular}{|ccc|}
\hline
$k_BT/\epsilon$ & $\langle U\rangle/(\epsilon n)$ & error \\
\hline
$\infty$&-0.888	&0\\
0.8&    -       &        -\\
0.6&	-1.12& 	0.03\\
0.5&	-1.39&	0.03\\
0.4&	-1.87&	0.02\\
0.3&	-2.38& 	0.01\\
0.2&	-2.85& 	0.02\\
0.1&	-3.1	&	$\approx$0\\
\hline 
\end{tabular}
\begin{tabular}{|ccc|}
\hline
$k_BT/\epsilon$ & $\langle U\rangle/(\epsilon n)$ & error \\
\hline
$\infty$&-0.9		&0\\
0.8	&-	&-\\
0.6	&-	&-\\
0.5&	-1.36&	0.04\\
0.4&	-1.88&	0.02\\
0.3&	-2.46&	0.02\\
0.2&	-2.94&	0.03\\
0.1&	-3.2	&	$\approx$0\\
\hline 
\end{tabular}
\begin{tabular}{|ccc|}
\hline
$k_BT/\epsilon$ & $\langle U\rangle/(\epsilon n)$ & error \\
\hline
$\infty$&-0.909	&0\\
0.8	&-	&-\\
0.6	&-	&-\\
0.5&	-1.35& 	0.03\\
0.4&	-1.96&	0.03\\
0.3&	-2.55& 	0.02\\
0.2&	-3.09& 	0.09\\
0.1&	-3.36	&	$\approx$0\\
\hline 
\end{tabular}
\begin{tabular}{|ccc|}
\hline
$k_BT/\epsilon$ & $\langle U\rangle/(\epsilon n)$ & error \\
\hline
$\infty$&-0.916	&0\\
0.8	&-	&-\\
0.6	&-	&-\\
0.5&	-1.28&	0.04\\
0.4&	-1.92&	0.04\\
0.3&	-2.57&	0.02\\
0.2&	-3.00&	0.02\\
0.1&	-3.42	&	$\approx$0\\
\hline 
\end{tabular}
\caption{The tables refer from left to right to clusters made up of
$n=3,4,\ldots,12$ particles. $U$ is the potential energy of a cluster of $n$
particles. Below $k_BT/\epsilon=0.1$ the reduced excess internal energy
per particle remains roughly constant in all cases: the smoothing
procedure described in Sec. \ref{sec:ic} was used. The data was
obtained with a Monte Carlo simulation over 5 million steps where one
step consists of $n$ particles moves. The strategy {\sl i.} described in
Sec. \ref{sec:ic} was used in the simulations.}
\label{tab:1}
\end{table}
\begin{table}[H]\tiny
\begin{tabular}{|ccc|}
\hline
$k_BT/\epsilon$ & $\langle U\rangle/(\epsilon n)$ & error \\
\hline
$\infty$&-0.666   	&0\\
0.8&	-0.7211&	0.0002\\
0.6&	-0.7437& 	0.0003\\
0.5&	-0.7659&  	0.0004\\
0.4&	-0.8052&	0.0005\\
0.3&	-0.8723& 	0.0007\\
0.2&	-0.9647&	0.0005\\
0.1&	-0.99881&	0.00005\\
\hline 
\end{tabular}
\begin{tabular}{|ccc|}
\hline
$k_BT/\epsilon$ & $\langle U\rangle/(\epsilon n)$ & error \\
\hline
$\infty$&-0.75   	&0\\
0.8&	-0.8466&	0.0005\\
0.6&	-0.8995& 	0.0009\\
0.5&	-0.959&  	0.001\\
0.4&	-1.073&	0.002\\
0.3&	-1.280& 	0.002\\
0.2&	-1.4597&	0.0009\\
0.1&	-1.49871&	0.00006\\
\hline 
\end{tabular}
\begin{tabular}{|ccc|}
\hline
$k_BT/\epsilon$ & $\langle U\rangle/(\epsilon n)$ & error \\
\hline
$\infty$&-0.9		&0\\
0.8&	-1.066&	0.001\\
0.6&	-1.200&	0.003\\
0.5&	-1.418&	0.009\\
0.4&	-1.884&	0.009\\
0.3&	-2.46&	0.01\\
0.2&	-2.96&	0.03\\
0.1&	-3.1982&	0.0006\\
\hline 
\end{tabular}
\begin{tabular}{|ccc|}
\hline
$k_BT/\epsilon$ & $\langle U\rangle/(\epsilon n)$ & error \\
\hline
$\infty$&-0.909	&0\\
0.8&	-1.078&	0.002\\
0.6&	-1.215&	0.003\\
0.5&	-1.423& 	0.008\\
0.4&	-1.90&	0.01\\
0.3&	-2.52& 	0.02\\
0.2&	-3.13& 	0.04\\
0.1&	-3.16&	0.01\\
\hline 
\end{tabular}
\caption{The tables refer, from left to right, to clusters made up of
$n=3,4,10,11$ particles. $U$ is the potential energy of a cluster of $n$
particles. The smoothing procedure described in Sec. \ref{sec:ic}
was not used at the lowest temperature. The strategy {\sl ii.} described
in Sec. \ref{sec:ic} was used in the simulations.}
\label{tab:1.5}
\end{table}
\begin{table}[H]\tiny
\begin{tabular}{|ccc|}
\hline
$k_BT/\epsilon$ & $\langle U\rangle/(\epsilon n)$ & error \\
\hline
$\infty$&-0.666   	&0\\
0.7&	-0.705& 	0.002\\
0.5&	-0.732&  	0.002\\
0.3&	-0.832& 	0.002\\
0.1&	-0.99872&	0.00008\\
\hline 
\end{tabular}
\begin{tabular}{|ccc|}
\hline
$k_BT/\epsilon$ & $\langle U\rangle/(\epsilon n)$ & error \\
\hline
$\infty$&-0.75   	&0\\
0.7&	-&  	-\\
0.5&	-0.866&  	0.007\\
0.3&	-1.138&	0.005\\
0.1&	-1.4987&	0.0002\\
\hline 
\end{tabular}
\begin{tabular}{|ccc|}
\hline
$k_BT/\epsilon$ & $\langle U\rangle/(\epsilon n)$ & error \\
\hline
$\infty$&-0.8   	&0\\
0.7&	-0.87&	0.02\\
0.5&	-1.00&	0.03\\
0.3&	-1.427& 	0.008\\
0.1&	-1.7984&	0.0002\\
\hline 
\end{tabular}
\begin{tabular}{|ccc|}
\hline
$k_BT/\epsilon$ & $\langle U\rangle/(\epsilon n)$ & error \\
\hline
$\infty$&-0.833	&0\\
0.7&	-&	-\\
0.5&	-0.95&	0.01\\
0.3&	-1.63&	0.01\\
0.1&	-2.1656&	0.0002\\
\hline 
\end{tabular}
\begin{tabular}{|ccc|}
\hline
$k_BT/\epsilon$ & $\langle U\rangle/(\epsilon n)$ & error \\
\hline
$\infty$&-0.857	&0\\
0.7&	-& 	-\\
0.5&	-0.95&	0.01\\
0.3&	-1.79&	0.01\\
0.1&	-2.22&	0.02\\
\hline 
\end{tabular}
\begin{tabular}{|ccc|}
\hline
$k_BT/\epsilon$ & $\langle U\rangle/(\epsilon n)$ & error \\
\hline
$\infty$&-0.875		&0\\
0.7&	-&	-\\
0.5&	-&	-\\
0.3&	-1.91&	0.03\\
0.1&	-2.3706&	0.0009\\
\hline 
\end{tabular}
\begin{tabular}{|ccc|}
\hline
$k_BT/\epsilon$ & $\langle U\rangle/(\epsilon n)$ & error \\
\hline
$\infty$&-0.888	&0\\
0.7&    -       &        -\\
0.5&	-&	 -\\
0.3&	-1.95& 	0.02\\
0.1&	-2.4416&	0.0005\\
\hline 
\end{tabular}
\begin{tabular}{|ccc|}
\hline
$k_BT/\epsilon$ & $\langle U\rangle/(\epsilon n)$ & error \\
\hline
$\infty$&-0.9		&0\\
0.7	&-	&-\\
0.5&	-&	-\\
0.3&	-2.07&	0.04\\
0.1&	-2.5969&	0.0006\\
\hline 
\end{tabular}
\begin{tabular}{|ccc|}
\hline
$k_BT/\epsilon$ & $\langle U\rangle/(\epsilon n)$ & error \\
\hline
$\infty$&-0.909	&0\\
0.7	&-	&-\\
0.5&	-& 	-\\
0.3&	-2.10& 	0.04\\
0.1&	-2.721&	0.002\\
\hline 
\end{tabular}
\begin{tabular}{|ccc|}
\hline
$k_BT/\epsilon$ & $\langle U\rangle/(\epsilon n)$ & error \\
\hline
$\infty$&-0.916	&0\\
0.7	&-	&-\\
0.5&	-&	-\\
0.3&	-2.01&	0.03\\
0.1&	-2.730&	0.008\\
\hline 
\end{tabular}
\caption{Same as Table \ref{tab:1} but with $\Delta=0.25\sigma$. The
  smoothing procedure described in Sec. \ref{sec:ic} was not used at
  the lowest temperature.} 
\label{tab:4}
\end{table}
\clearpage
\begin{table}[H]\tiny
\begin{tabular}{|ccc|}
\hline
$k_BT/\epsilon$ & $\langle U\rangle/(\epsilon n)$ & error \\
\hline
$\infty$&-0.666   	&0\\
0.7& 	-0.6914&	0.0003\\
0.5&	-0.7114&	0.0004\\
0.3&	-0.792& 	0.001\\
0.1&	-0.9987&	0.0002\\
\hline 
\end{tabular}
\begin{tabular}{|ccc|}
\hline
$k_BT/\epsilon$ & $\langle U\rangle/(\epsilon n)$ & error \\
\hline
$\infty$&-0.75   	&0\\
0.7&	-0.7903&	0.0007\\
0.5&	-0.826&  	0.002\\
0.3&	-1.138&	0.005\\
0.1&	-1.49871&	0.00006\\
\hline 
\end{tabular}
\begin{tabular}{|ccc|}
\hline
$k_BT/\epsilon$ & $\langle U\rangle/(\epsilon n)$ & error \\
\hline
$\infty$&-0.8   	&0\\
0.7&	-0.8473&	0.0009\\
0.5&	-0.895&	0.002\\
0.3&	-1.230&	0.008\\
0.1&	-1.7989&	0.0001\\
\hline 
\end{tabular}
\begin{tabular}{|ccc|}
\hline
$k_BT/\epsilon$ & $\langle U\rangle/(\epsilon n)$ & error \\
\hline
$\infty$&-0.833	&0\\
0.7&	-0.884&	0.001\\
0.5&	-0.936&	0.002\\
0.3&	-1.35&	0.01\\
0.1&	-1.9985&	0.0004\\
\hline 
\end{tabular}
\begin{tabular}{|ccc|}
\hline
$k_BT/\epsilon$ & $\langle U\rangle/(\epsilon n)$ & error \\
\hline
$\infty$&-0.857	&0\\
0.7&	-0.913&	0.001\\
0.5&	-0.955&	0.002\\
0.3&	-1.61&	0.03\\
0.1&	-2.2848&	0.0001\\
\hline 
\end{tabular}
\begin{tabular}{|ccc|}
\hline
$k_BT/\epsilon$ & $\langle U\rangle/(\epsilon n)$ & error \\
\hline
$\infty$&-0.875		&0\\
0.7&	-0.928&	0.001\\
0.5&	-0.980&	0.003\\
0.3&	-1.63&	0.03\\
0.1&	-2.371&	0.001\\
\hline 
\end{tabular}
\begin{tabular}{|ccc|}
\hline
$k_BT/\epsilon$ & $\langle U\rangle/(\epsilon n)$ & error \\
\hline
$\infty$&-0.888	&0\\
0.7&	-0.945&	0.001\\
0.5&	-1.000&	0.003\\
0.3&	-1.55&	0.06\\
0.1&	-2.51&	0.04\\
\hline 
\end{tabular}
\begin{tabular}{|ccc|}
\hline
$k_BT/\epsilon$ & $\langle U\rangle/(\epsilon n)$ & error \\
\hline
$\infty$&-0.9		&0\\
0.7&	-0.956&	0.001\\
0.5&	-1.013&	0.004\\
0.3&	-1.56&	0.05\\
0.1&	-2.396&	0.001\\
\hline 
\end{tabular}
\begin{tabular}{|ccc|}
\hline
$k_BT/\epsilon$ & $\langle U\rangle/(\epsilon n)$ & error \\
\hline
$\infty$&-0.909	&0\\
0.7&	-0.9655&	0.0009\\
0.5&	-1.022&	0.004\\
0.3&	-1.61& 	0.03\\
0.1&	-2.5427&	0.0004\\
\hline 
\end{tabular}
\begin{tabular}{|ccc|}
\hline
$k_BT/\epsilon$ & $\langle U\rangle/(\epsilon n)$ & error \\
\hline
$\infty$&-0.916	&0\\
0.7&	-0.973&	0.001\\
0.5&	-1.033&	0.002\\
0.3&	-1.59&	0.02\\
0.1& 	-2.66&	0.004\\
\hline 
\end{tabular}
\caption{Same as Table \ref{tab:1} but with $\Delta=0.15\sigma$. The
  smoothing procedure described in Sec. \ref{sec:ic} was not used at
  the lowest temperature. The strategy {\sl ii.} described in
  Sec. \ref{sec:ic} was used in the simulation.} 
\label{tab:6}
\end{table}

\begin{table}[H]\scriptsize
\begin{center}
\begin{tabular}{|lll||ll||ll||l|}
\hline
& $\Delta=0.5\sigma$ && $\Delta=0.25\sigma$ && $\Delta=0.15\sigma$&&\\
\hline
$n$ & $a_n$ & $b_n$ &  $a_n$ & $b_n$ & $a_n$ & $b_n$ & $c_n=-(n-1)/n$\\
\hline
2&         0&         1&       	0&         1&        0&         1&        -0.5\\
3&        -0.337525&  3.88039&  -0.33890&  6.9050 &  -0.345587&  10.7799&  -0.66666\\
4&        -0.778556&  4.66976&  -0.77059&  7.5017 &  -0.773523&  7.97531&  -0.75\\
5&        -1.22587&   5.16189&  -1.0248 &  5.8901 &  -1.03428&   9.36621&  -0.8\\
6&        -1.69844&   5.59919&  -1.3810 &  7.3613 &  -1.20676&   9.21365&  -0.83333\\
7&        -1.89814&   5.26287&  -1.4235 &  6.7666 &  -1.47964&   8.27638&  -0.85714\\
8&	  -2.06452&   5.07916&  -1.5201 &  4.1792 &  -1.55091&   8.50313& -0.875\\
9&        -2.30070&   5.47737&  -1.5793 &  4.3672 &  -1.68144&   10.1592&  -0.88888\\
10&	  -2.39363&   5.50909&  -1.7253 &  4.2708 &  -1.55096&   9.41914& -0.9\\
11&       -2.55636&   5.64409&  -1.8464 &  4.8294 &  -1.69591&   9.75528&  -0.90909\\
12&	  -2.59747&   6.07744&  -1.8541 &  5.7234 &  -1.81374&
10.5661& -0.91666\\
\hline
\end{tabular}
\end{center}
\caption{Fit to the Gaussian of Eq. (\ref{ugauss}) of the reduced
  excess internal energy per particle of the first eleven $n$-clusters as a
  function of temperature.}  
\label{tab:fit}
\end{table}
%
\begin{acknowledgments}
I would like to acknowledge the support of the National
Institute for Theoretical Physics of South Africa.
\end{acknowledgments}

\bibliography{janus}

\begin{thebibliography}{39}
\bibitem [{sta()}]{stat}%
  \BibitemOpen
  \href@noop {} {}\bibinfo {note} {The quantum effects start playing a role
  when the de Broglie thermal wavelength $\Lambda=\sqrt{2\pi\hbar^2/(k_BTm)}$
  becomes comparable to the particle diameter $\sigma$. At room temperature
  this means that the nanoparticles should have a mass of the order of
  $10^{-26}Kg$ whereas the microparticles should have a mass of the order of
  $10^{-32}Kg$.}\BibitemShut {Stop}%
\end{thebibliography}
\end{document}